\documentstyle[twoside,fleqn,espcrc2,epsfig]{article}

\newcommand{\AmS}{{\protect\the\textfont2
  A\kern-.1667em\lower.5ex\hbox{M}\kern-.125emS}}

\font\sfhuge= cmss24  at 20truept
\font\sfLARGE= cmss14  at 14truept
  at 12truept
\font\sfmed= cmss10  at 10truept
\font\sfsml= cmss8   at  8truept

\title{Combined QCD analysis of e$^+$e$^-$ data at 
             $\sqrt{s} = 14$ to $172$~GeV 
       }

\author{O. Biebel\address{III. Physikalisches Institut A,
        RWTH Aachen --- Physikzentrum, \\ 
        Sommerfeldstr., D-53056 Aachen, Germany}%
       }
       
\begin{document}
\addtolength{\textheight}{-68mm}
%
\begin{titlepage}
\thispagestyle{empty}
\vspace*{-10mm}
\vbox to 245mm{ 
\hbox to \textwidth{ \hsize=\textwidth
\vbox{
\hbox {
\epsfig{file=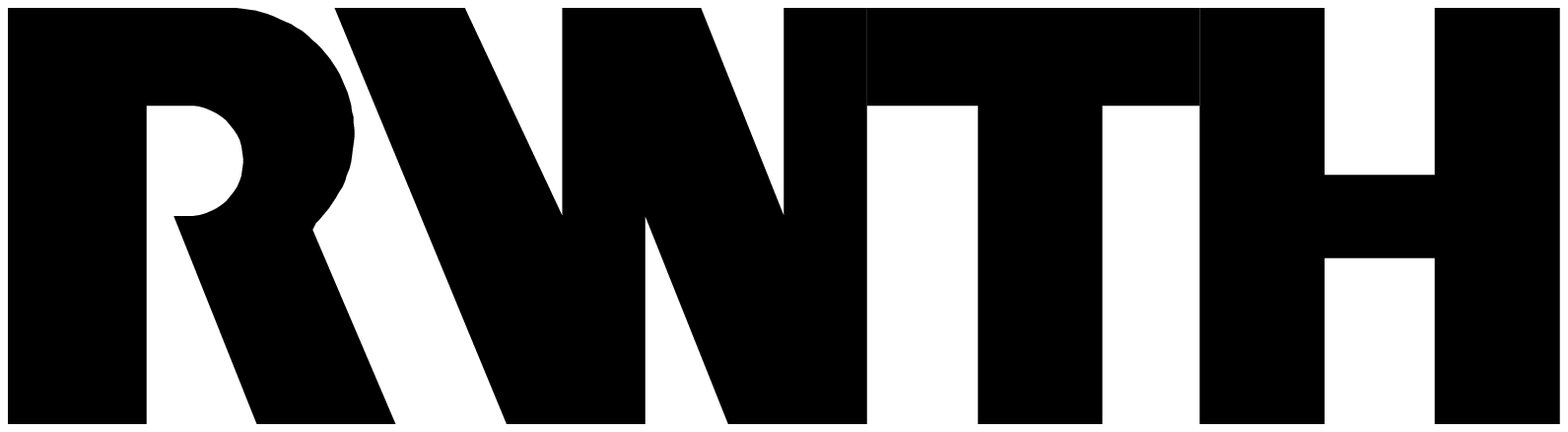,height=20mm}
} 
}
\vbox{
{
\hbox{\sfmed RHEINISCH-\hss}\vspace*{+0.150mm}
\hbox{\sfmed WESTF\"ALISCHE-\hss}\vspace*{+0.150mm}
\hbox{\sfmed TECHNISCHE-\hss}\vspace*{+0.150mm}
\hbox{\sfmed HOCHSCHULE-\hss}\vspace*{+0.150mm}
\hbox{\sfmed AACHEN\hss}\vspace*{0.200mm}
}
}
\vbox{ \hsize=58mm 
{
\hspace*{0pt\hfill}\hbox{\sfLARGE\hspace*{0pt\hfill}        PITHA 97/32\hss}\vspace*{0mm}
\hspace*{0pt\hfill}\hbox{        \hspace*{0pt\hfill} \rule{45mm}{1.0mm}\hss}\vspace*{2mm}
\hspace*{0pt\hfill}\hbox{\sfLARGE\hspace*{0pt\hfill}       August 1997\hss}
}
}
}

\vspace*{5cm}
\begin{center}
{\huge\bf
Combined QCD analysis of e$\mathbf{^+}$e$\mathbf{^-}$ data \\[2mm]
  at $\mathbf{\sqrt{s}}$ = 14 to 172 GeV
}
\end{center}
\vspace*{2cm}
\begin{center}
\Large
Otmar~Biebel \\
\bigskip 
\bigskip
III. Physikalisches Institut, Technische Hochschule Aachen\\
D-52056 Aachen, Germany
\end{center}

\vspace*{0pt\vfill}
\vfill

\vspace*{-5mm}
\noindent
\hspace*{-5mm}
\hbox {
\rule{\textwidth}{0.3mm}
}

\vspace*{6mm}
\noindent
\begin{minipage}{\textwidth}
\vbox {\vsize=60mm
\hbox to \textwidth{\hsize=\wd0
\hbox {\hspace*{-5mm}
\vbox{ 
\hbox to \textwidth{\hss\sfhuge PHYSIKALISCHE INSTITUTE\hss }\vspace*{3.0mm}
\hbox to \textwidth{\hss\sfhuge      RWTH AACHEN\hss }\vspace*{3.0mm}
\hbox to \textwidth{\hss\sfhuge D-52056 AACHEN, GERMANY\hss}
}
}
}
}
\end{minipage}
}
\end{titlepage}
\setlength{\textheight}{202mm}
\setlength{\textwidth}{160mm}
\setlength{\oddsidemargin}{-4mm}
\setlength{\evensidemargin}{4mm}
\setlength{\topmargin}{16mm}
\setlength{\headheight}{13mm}
\setlength{\headsep}{21pt}
\setlength{\footskip}{30pt}
\clearpage
\begin{abstract}
\vspace*{-46mm}
\hbox{\sfsml Talk presented at the {\em QCD'97}, 
             Montpellier, France, July 3-9, 1997.
}
\vspace*{43mm}
A study of the energy dependence of event shape observables is presented. 
The strong coupling constant $\alpha_s$ has been determined from the mean 
values of six event shape observables. Power corrections, employed for
the measurement of $\alpha_s$, have been found to approximately account 
for hadronisation effects.
\end{abstract}

\maketitle

\section{INTRODUCTION}
Many years of electron-positron annihilation experiments have yielded 
a huge amount of data covering a vast region of centre-of-mass energies. 
For example, at the PEP, PETRA, TRISTAN and LEP colliders
annihilation experiments have been performed between $\sqrt{s} = 14$ and 
$172$~GeV~(see e.g. \cite{bib-bethke} and references therein).
Distributions of event shape observables measured from hadronic
final states clearly show a dependence on the centre-of-mass 
energy~(e.g. \cite{bib-fernandez}). 
This violation of the scaling behaviour is due to the running of 
$\alpha_s(\mu)$ with the energy scale $\mu = x_{\mu} \cdot \sqrt{s}$.

Every measurement of the value of $\alpha_s$ from event shape 
distributions requires the treatment of impacts of the hadronisation 
of quarks and gluons which cannot be calculated perturbatively. Usually,
phenomenological hadronisation models, available from Monte Carlo event 
generators, are employed to correct for hadronisation effects. An approach, 
independent of hadronisation models, to account for such non-perturbative
effects are power corrections \cite{bib-dokshitzer,bib-webber}. 
In the following, power corrections to the mean values of 
event shape dis\-tributions are investigated in order to 
determine the strong coupling constant $\alpha_s$ from the scaling
violation that is seen in the energy dependence of the mean values.

\section{POWER CORRECTIONS}
Non-perturbative contributions to event shape distributions are, on general 
grounds, expected to give rise to corrections proportional to reciprocal 
powers of $\sqrt{s}$. Hadronisation effects, for example, should yield a 
$1/\sqrt{s}$ correction according to the {\em tube model}~\cite{bib-tube}
in which hadrons 
emanate from a tube spanned in rapidity and transverse momentum space 
between quark and anti-quark. Another, more formal, theoretical approach 
leading to power corrections deals with renormalons~\cite{bib-renormalon},
that are found in 
the QCD vacuum polarisation at high order perturbative calculations. Power 
corrections may also be explained in a more phenomenological ansatz 
\cite{bib-dokshitzer} with the behaviour of the strong coupling at low scales 
$\mu \ll 1$~GeV.

The perturbative expression for the energy dependence of $\alpha_s(\mu)$
gives rise to the unphysical Landau pole at $\mu=\Lambda_{\mathrm{QCD}}$. 
In general, one would, in this energy regime, expect the running such that 
$\alpha_s(\mu)$ remains finite but eventually large for all scales down 
to zero (see e.g. \cite{bib-shirkov}). In the approach of \cite{bib-dokshitzer}
an effective coupling constant
\begin{displaymath}
 \overline{\alpha}_{p-1}(\mu_I) \equiv \frac{p}{\mu_I^p}\int_0^{\mu_I}
                                \frac{{\mathrm{d}}\mu}{\mu}\ \alpha_s(\mu) \mu^p
\end{displaymath}
is introduced
that absorbs all non-per\-tur\-ba\-tive details of
$\alpha_s(\mu)$ up to an arbitrary infrared matching scale $\mu_I \simeq$~GeV.
Using the effective coupling, power corrections to, for example, the mean values
of event shapes can be calculated. They turn out to be simply additive
to the perturbative calculation 
$\langle{\cal F}\rangle = \langle{\cal F}_{\mathrm{pert.}}\rangle
                        + \langle{\cal F}_{\mathrm{pow.}}\rangle$. 
The principal structure of the power correction is~\cite{bib-dokshitzer}
\begin{eqnarray*}
\langle{\cal F}_{\mathrm{pow.}}\rangle 
                        & \sim & a_{\cal F} \cdot
        \left(\frac{\mu_I}{\sqrt{s}}\right)^p \cdot
        \ln ^r\left(\frac{\sqrt{s}}{\mu_I}\right) \cdot \nonumber \\
                        &      &  \cdot
        \left[ \overline{\alpha}_{p-1}(\mu_I) -
         \{\alpha_s(\mu) + \ldots \} \right]
\end{eqnarray*}
in which  $a_{\cal F}$ is a coefficient, $p=1$, $2$ is the power of the 
$1/\sqrt{s}$ 
term, $r=0$, $1$ is the power of an enhanced power term. All these parameters
depend 
on the particular event shape. The term in curly brackets accounts for the 
contribution of the perturbative expression for the running of $\alpha_s(\mu)$ 
at $\mu < \mu_I$. It is know in next-to-leading order (${\cal O}(\alpha_s^2)$).

The coefficients and parameters of the power corrections have been calculated 
for many event shape observables~(e.g. \cite{bib-dokshitzer,bib-webber}), 
in particular, for mean values of thrust $T$, $C$ parameter, heavy jet mass 
$M_H^2/s$, total and wide jet broadening (respectively, $B_T$ and $B_W$) and 
the differential 2-jet rate $y_{23}$ using the Durham scheme (cf. 
Table~\ref{tab-powcor-params})
\begin{table}
\caption{\label{tab-powcor-params}
Type of power correction to mean values of event shapes. 
`?' indicates an unknown coefficient.
}
\renewcommand{\arraystretch}{1.4}
\begin{tabular}{|c||c|c|c|c|}
\hline
observable ${\cal F}$   &     $a_{\cal F}$      &    power terms \\
\hline
$\langle T\rangle$      &       $-1$            &    $1/\sqrt{s}$  \\
$\langle C\rangle$      &       $3\pi/2$        &    $1/\sqrt{s}$  \\
$\langle M_H^2/s\rangle$&       $1.0 \pm 0.5$   &    $1/\sqrt{s}$  \\
\hline
$\langle B_T\rangle$    &       $1.0 \pm 0.5$   &    
                                     $\ln(\sqrt{s})/\sqrt{s} + ?\cdot1/\sqrt{s}$ \\
$\langle B_W\rangle$    &       $1.0 \pm 0.5$   &    
                                     $\ln(\sqrt{s})/\sqrt{s} + ?\cdot1/\sqrt{s}$ \\
\hline
$\langle y_{23}\rangle$ &       ?               &    
                                               $ ?\cdot1/s + ?\cdot\ln(\sqrt{s})/s$ \\
\hline
\end{tabular}
\end{table}

\section{TEST OF POWER CORRECTIONS}
\label{sec-test}
\begin{figure}
\epsfig{file=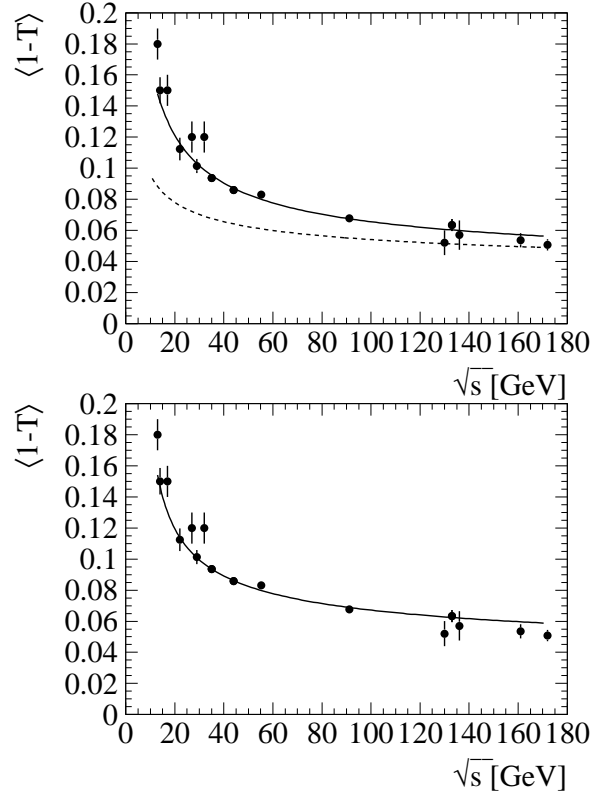,width=0.5\textwidth}
\vspace*{-15mm}
\caption{\label{fig-thrust-xmu}
Fits to thrust. Top: The solid line is the result of
a fit of $\overline{\alpha}_{0}$ with 
$\alpha_s$ and $x_{\mu}$ fixed. The dashed line is the 
perturbative prediction only.
Bottom: Fit of $\alpha_s$ and $x_{\mu}$ without power corrections.
}
\end{figure}

A comparison of the perturbative prediction with 
power corrections with plain perturbative predictions
has been performed. Figure~\ref{fig-thrust-xmu} shows the results of two
fits for the thrust. 
The first fit including the dominating $1/\sqrt{s}$ power correction term 
obtained $\chi^2/{\mathrm{d.o.f}}= 1.8$ for
$\overline{\alpha}_{0}(\mu_I=2~{\mathrm{GeV}})=0.57 \pm 0.01$ with fixed 
values for $\alpha_s \equiv 0.122$ and $x_{\mu} \equiv 1$,\footnote{If not 
stated explicitly, $\alpha_s$ refers to $\alpha_s(M_{\mathrm{Z}^0})$.}
which is 
the LEP average from event shapes~\cite{bib-bethke}. The difference to the
dashed line in Figure~\ref{fig-thrust-xmu} reveals the size of the power 
corrections. The 
second fit, using the perturbative prediction only, resulted in  
$\chi^2/{\mathrm{d.o.f}}= 2.4$ for
$\alpha_s=0.122\pm 0.001$ and $x_{\mu} = 0.07 \pm 0.01$. The inclusion of 
power corrections, thererfore, improves slighly the description of the data. 
Thus it allows to determine $\alpha_s$ with\-out the need for a very small
renormalisation scale $\mu = x_{\mu} \cdot \sqrt{s}$ which approximates 
hadronisation effects.

\begin{figure}
\epsfig{file=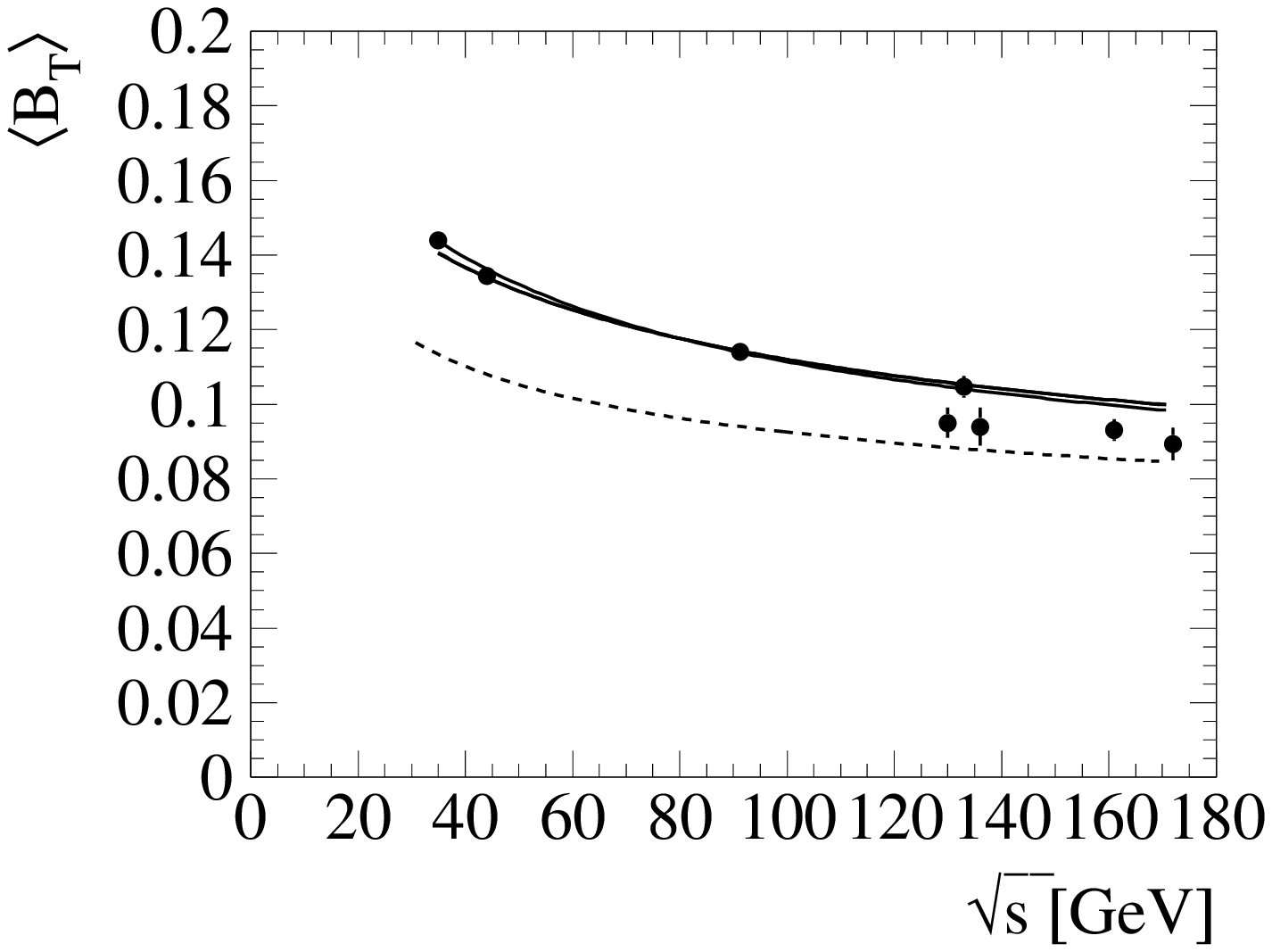,width=0.5\textwidth}
\epsfig{file=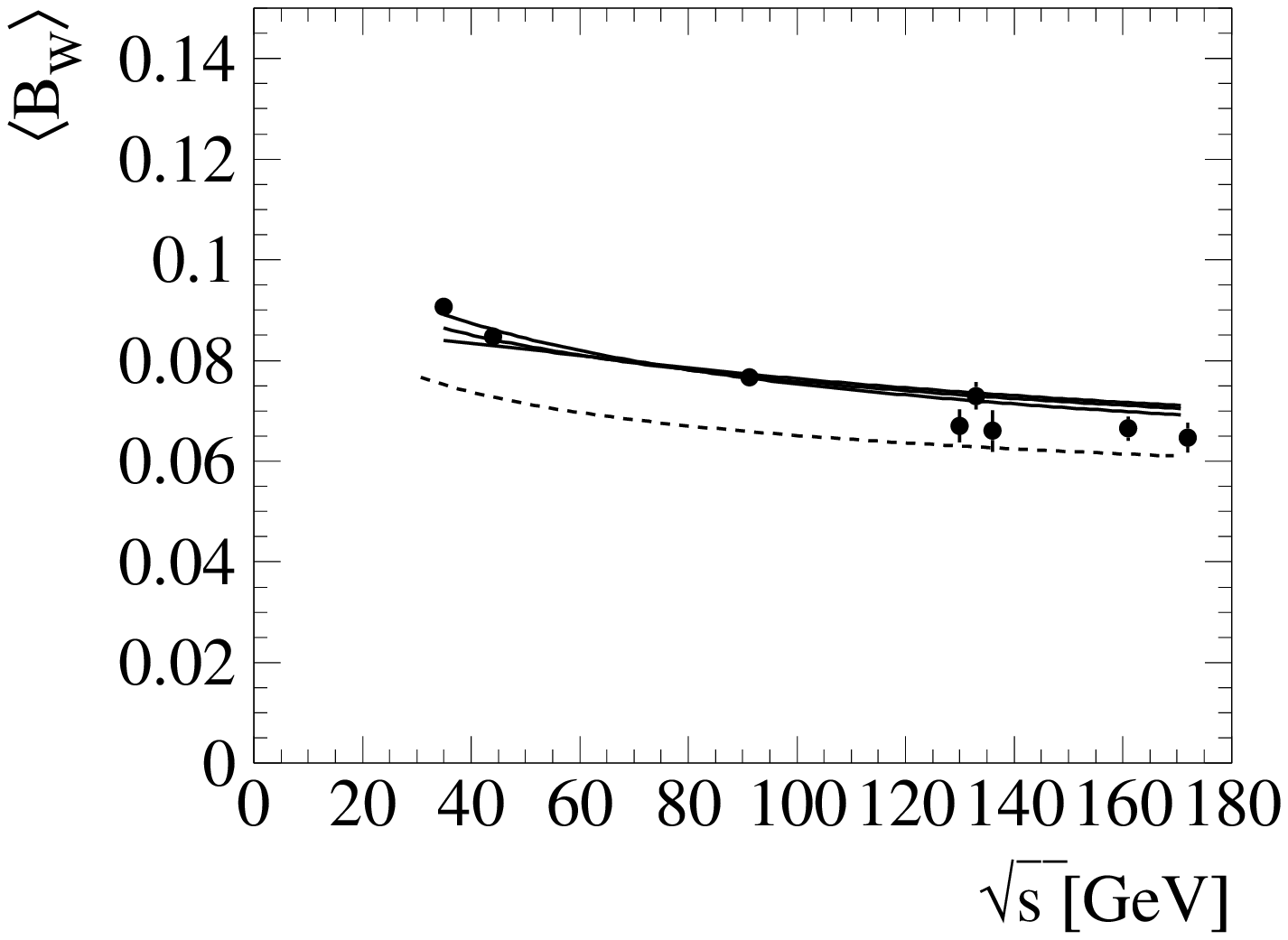,width=0.5\textwidth}
\vspace*{-15mm}
\caption{\label{fig-bt-bw}
Fits to total (top) and wide (bottom) jet broadening: The solid lines are 
the results of the fits for $\overline{\alpha}_{0}$ with $\alpha_s$ and 
$x_{\mu}$ fixed. The dashed lines are the perturbative prediction only.
}
\end{figure}

The two jet broadening measures are expected to suffer enhanced power terms. 
Additionally, non-enhanced corrections are expected but the coefficients are 
difficult to predict~\cite{bib-webber}. Due to this, fits have been performed
with ``enhanced only'' power corrections, ``non-enhanced only'' and a 
combination of 
both. Using fixed values for $\alpha_s$ and $x_{\mu}$ as before, the fits
obtained $\chi^2/{\mathrm{d.o.f}}$ of $4.6$, $3.3$, $5.2$, respectively, for 
$\langle B_T\rangle$ and $11.3$, $2.5$, $5.5$, respectively, for
$\langle B_W\rangle$. Figure~\ref{fig-bt-bw} shows these results as
solid lines, whereas the dashed line represents the plain perturbative
prediction. A slight preference for ``non-enhanced only'' power corrections
might be deduced from the fit results for $\langle B_W\rangle$. 
However, the data are still not
sufficient to decide in favour of one of the three variations.

For $\langle y_{23}\rangle$ it is only known that the power correction 
should be quadratic in $1/\sqrt{s}$ including an enhanced 
term~\cite{bib-webber-privcomm}. Therefore, the $a_{y}$ coefficient
has been fitted for, too. Fits have been tried with enhanced and non-enhanced
power corrections of the type $1/\sqrt{s}$ and $1/s$. By fixing the
values of $\alpha_s$ and $x_{\mu}$ as before, all fits obtained 
$\chi^2/{\mathrm{d.o.f.}}$ of unity. The $a_{y}$ came out to be of the 
order $10^{-4}$ to $10^{-6}$ and
compatible with zero within the errors 
suggesting that power corrections are negligible for the mean
value of the differential 2-jet rate using the Durham scheme. 
\begin{figure}
\epsfig{file=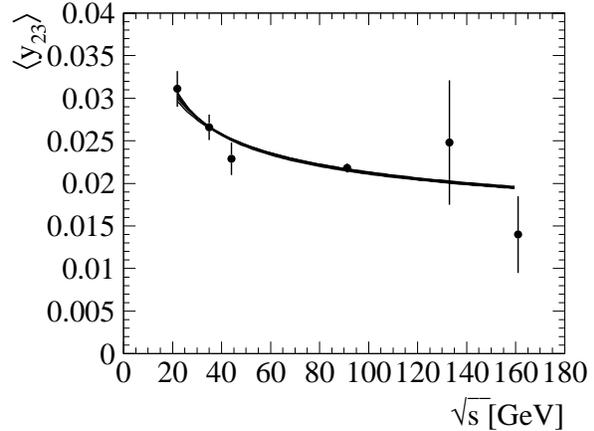,width=0.5\textwidth}
\vspace*{-15mm}
\caption{\label{fig-y23}
Fits to mean differential 2-jet rate: The solid lines are the results of the 
fits for $a_{y}$ and $\overline{\alpha}_{0}$ with $\alpha_s$ and $x_{\mu}$ 
fixed. 
}
\end{figure}

\section{DETERMINATION OF $\alpha_s$}
The value of $\alpha_s$ has been obtained by fitting the perturbative
calculation plus power corrections to the data using 
$\alpha_s(x_{\mu}\cdot M_{\mathrm{Z}^0})$
and 
$\overline{\alpha}_{0}(\mu_I=2~{\mathrm{GeV}})$ 
as parameters of the fit. 
Since the data still cannot yet decide in favour of non-enhanced power 
corrections for 
the jet broadening, the enhanced corrections predicted in \cite{bib-webber}
have been used. For the mean value of the differential 2-jet rate using the 
Durham scheme the fit has been performed without power corrections as is 
suggested by the results of Section~\ref{sec-test}.
\begin{table}[t]
\caption{\label{tab-a0-as-chi}
Preliminary results and $\overline{\chi^2}\equiv\chi^2/{\mathrm{d.o.f.}}$ of 
the fits. The errors represent experimental uncertainties, $x_{\mu}$, 
and $a_{\cal F}$ variation. For $\alpha_s$ a variation of $\mu_I$ is 
included.
}
\renewcommand{\arraystretch}{1.4}
\begin{tabular}{|c||c|c|c|}
\hline
${\cal F}$              & $\overline{\alpha}_{0}(2~{\mathrm{GeV}})$ 
                                       & $\alpha_s(M_{\mathrm{Z}^0})$ 
                                             & $\overline{\chi^2}$     \\
\hline
$\langle T\rangle$      
                        & $0.543 \pm 0.014$
                                       & $0.120 ^{+0.007}_{-0.004}$ 
                                             & $1.8$                          \\
$\langle C\rangle$      
                        & $0.474 \pm 0.015$
                                       & $0.120 ^{+0.004}_{-0.003}$ 
                                             & $2.0$                          \\
$\langle M_H^2/s\rangle$
                        & $0.457 ^{+0.212}_{-0.077}$
                                       & $0.112 ^{+0.006}_{-0.005}$ 
                                             & $0.8$                          \\
\hline
$\langle B_T\rangle$    
                        & $0.342 ^{+0.064}_{-0.036}$
                                       & $0.116 ^{+0.007}_{-0.006}$ 
                                             & $4.2$                          \\
$\langle B_W\rangle$    
                        & $0.264 ^{+0.048}_{-0.031}$
                                       & $0.111 ^{+0.006}_{-0.004}$ 
                                             & $2.3$                          \\
\hline
$\langle y_{23}\rangle$ 
                        & ---
                                       & $0.123 ^{+0.006}_{-0.004}$ 
                                             & $0.9$                          \\
\hline
\end{tabular}
\end{table}

\begin{figure}
\epsfig{file=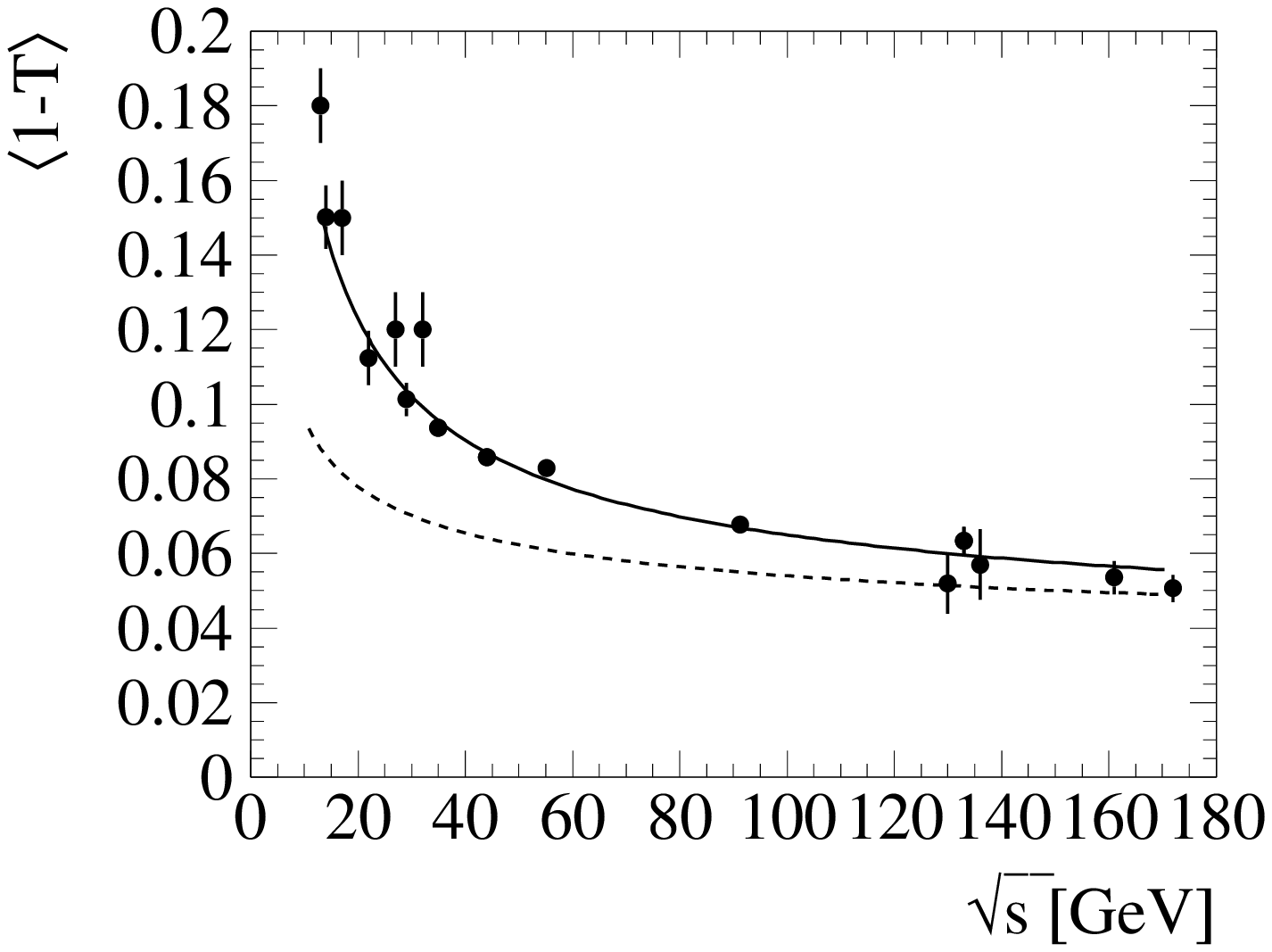,width=0.5\textwidth}
\epsfig{file=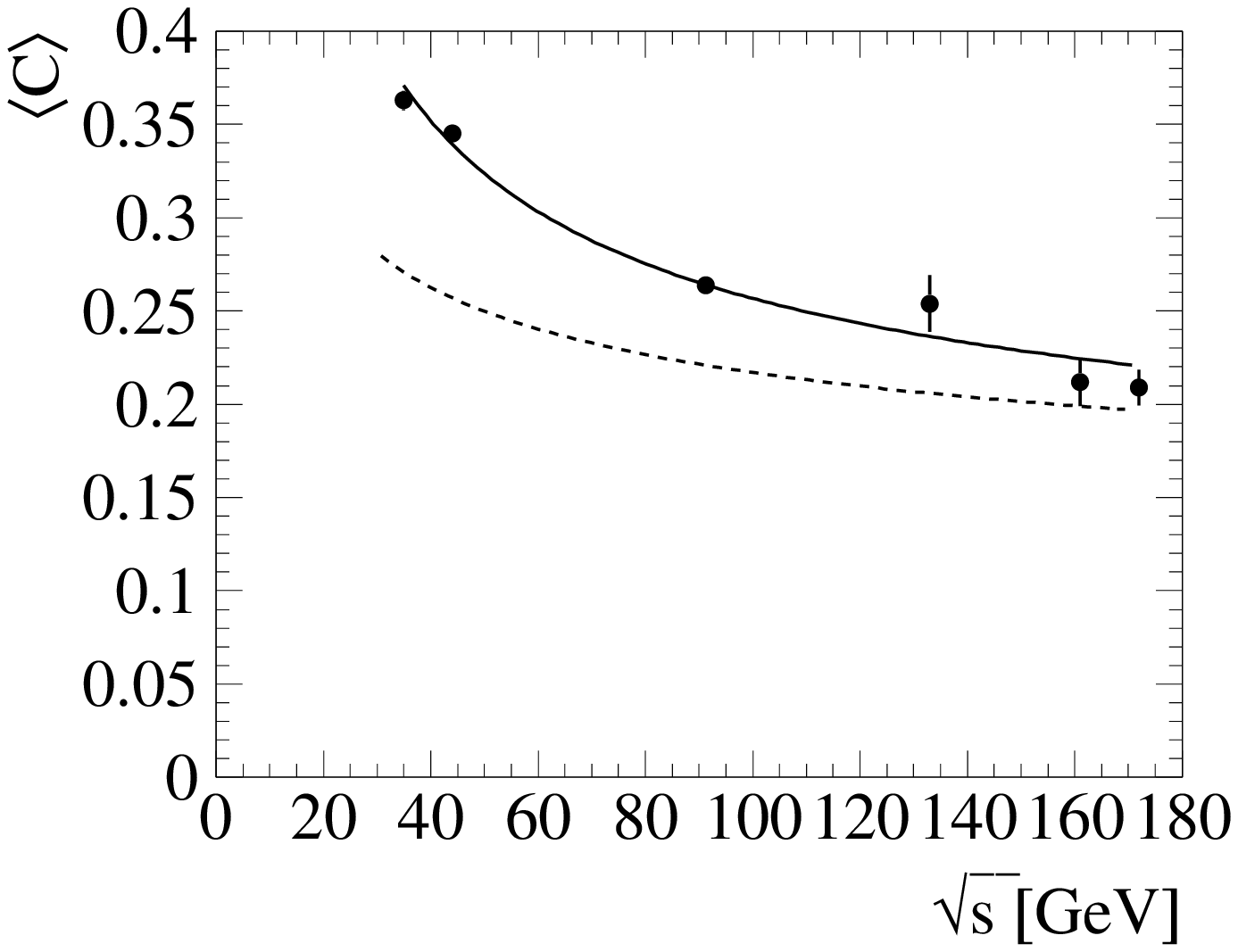,width=0.5\textwidth}
\epsfig{file=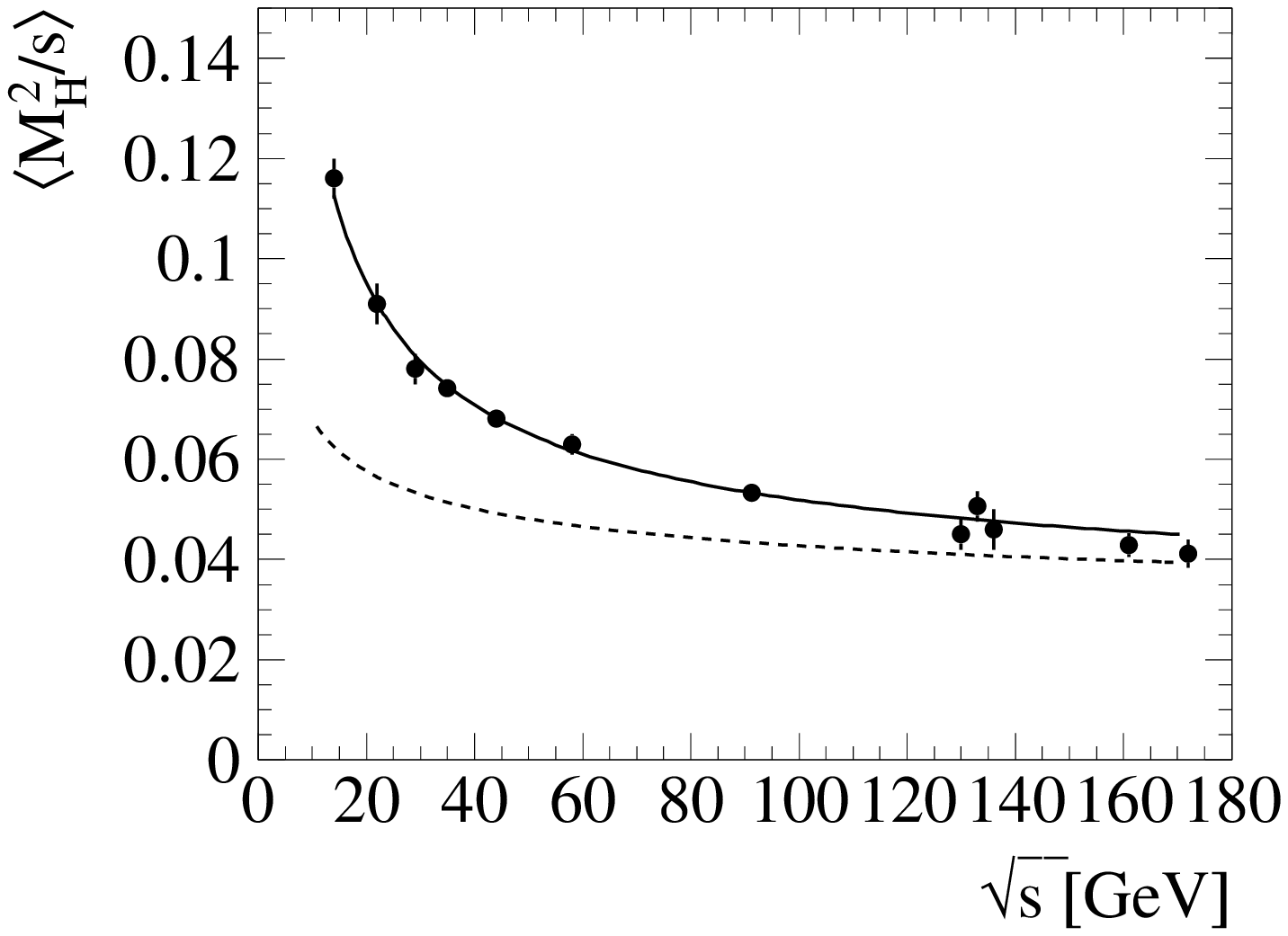,width=0.5\textwidth}
\vspace*{-15mm}
\caption{\label{fig-t-c-mh}
Fits to thrust (top), $C$ parameter (middle) and heavy jet mass (bottom): 
The solid lines are the results of the fits for $\overline{\alpha}_{0}$ 
and $\alpha_s$ with $x_{\mu}$ fixed. The dashed line is the perturbative 
prediction for the fitted $\alpha_s$ only.
}
\end{figure}

Figure~\ref{fig-t-c-mh} exemplifies the results
of such fits for thrust, $C$ parameter and heavy jet mass. 
For comparison the plain perturbative prediction using the fitted value of 
$\alpha_s$ is also shown by the dashed lines. For the central values of 
$\overline{\alpha}_{0}$ and $\alpha_s$, $x_{\mu}=1$ and $\mu_I=2$~GeV are 
used. The systematic uncertainties for $\overline{\alpha}_{0}$ are estimated 
by varying $x_{\mu}$ between $1/2$ and $2$ and by varying $a_{\cal F}$ as 
stated in Table~\ref{tab-powcor-params}. In addition, the errors of 
$\alpha_s$ include a variation of $\mu_I$ by $\pm 1$~GeV. The results 
of the fits and the $\chi^2/{\mathrm{d.o.f}}$ are 
given in Table~\ref{tab-a0-as-chi}.

Averaging the six values of $\alpha_s$ weighted by their total errors,
and considering correlations for the errors, yields
\begin{displaymath}
  \alpha_s(M_{\mathrm{Z}^0}) = 0.116 \pm 0.001 ^{+0.006}_{-0.004}
\end{displaymath}
where the first error is the experimental uncertainty and the second represent 
the deviations due to the variation of $x_{\mu}$, $\mu_I$, and $a_{\cal F}$
added quadratically. This value agrees with the world average of
$\alpha_s^{\mathrm{w.a.}}(M_{\mathrm{Z}^0}) = 0.118 \pm 0.003$ 
\cite{bib-bethke,bib-schmelling}.

\section{SUMMARY}
Experiments at e$^+$e$^-$ colliders have measured distributions of event shape 
observables over a large centre-of-mass range of $\sqrt{s}= 14$-$172$~GeV. 
The distributions clearly show a scaling violation which is expected due to
the running of the strong coupling constant $\alpha_s(\mu)$.

Hadronisation effects contained in these distributions can be explained by
soft gluon phenomena. Their theoretical treatment yields corrections
proportional to powers of $1/\sqrt{s}$ in addition to the usual perturbative
prediction. Power corrections have been found to be useful approximations of
hadronisation effects. They have been applied
to determine
$\alpha_s(M_{\mathrm{Z}^0}) = 0.116 ^{+0.006}_{-0.004}$
from the energy dependence of the mean values
of thrust, $C$ parameter, heavy jet mass, total and wide jet broadening, and 
of the differential 2-jet rate using the Durham scheme. 
The study also uncovered that the predicted power corrections are not universal 
for the event shape distributions considered. Thus, hadronisation effects cannot 
be explained completely by power behaved corrections.

\section*{QUESTIONS}
\begin{raggedright}
{\em L.~Trentadue, Parma:} \\
\end{raggedright}
I repeat here, too, the comment I have already made to the previous
talk by P.A. Movilla Fern\'andez. You have not considered in your
analysis an event shape variable i.e. the energy-energy correlation
for which the $1/\sqrt{s}$ corrections have for the first time
been observed. Such a quantity has same peculiar characteristics:
a) It involves experimentally and theoretically a single hadron
distribution so that hadronisation can be seen as a single hadron
mechanism instead as a ``collective'' effect; b) hadronisation
can be discussed and included analytically without using a Monte
Carlo; c) a complete next-to-leading resummed expression is also
known. It would be interesting in my opinion to see the results
of your analysis for such a variable too.

\vspace*{1ex}
\begin{raggedright}
{\em O.~Biebel:} \\
\end{raggedright}
I know about energy-energy correlations which have been investigated
already by several experiments. For this study I focussed on 
new observables for which measurements became available by the
work of P.A. Movilla Fern\'andez et al.~\cite{bib-fernandez}. 
Additionally, 
next-to-leading resummed expressions are known for all of the 
observables considered in this study. I should point out, that I 
used the ${\cal O}(\alpha_s^2)$ perturbative predictions only.

\vspace*{1ex}
\begin{raggedright}
{\em G.~Dissertori, CERN:} \\
\end{raggedright}
$\overline{\alpha}_{0}$ is a universal parameter, so it should be
the same for all observables. From your analysis it looks like this
is not the case. Can you comment on this?

\vspace*{1ex}
\begin{raggedright}
{\em O.~Biebel:} \\
\end{raggedright}
Indeed, a universal $\overline{\alpha}_0$ is incompatible with the data.
I forced $\overline{\alpha}_{0}$ to be universal in a combined fit of 
thrust, $C$ parameter and heavy jet mass which gave rather comparable
values for $\overline{\alpha}_{0}$ from the individual fits. The 
combined fit for these three yielded a huge $\chi^2/{\mathrm{d.o.f.}}$
and also the fit curve failed to describe the data. 

\vspace*{1ex}
\begin{raggedright}
{\em G.~Dissertori, CERN:} \\
\end{raggedright}
There is a new prediction for power-law corrections to the
distribution of thrust. Have you looked at the distributions,
too?

\vspace*{1ex}
\begin{raggedright}
{\em O.~Biebel:} \\
\end{raggedright}
I have not yet tried this. However, according
to Yu.L. Dokshitzer and B.R. Webber, the authors of this 
prediction~\cite{bib-dokshitzer-2}, 
the calculation is possible only for thrust and cannot be transferred to
other observables. They have fit their prediction in the range 
$0.05 < 1-T < 0.35$ to thrust distributions measured at 
$\sqrt{s} = 14$-$161$~GeV 
obtaining a sensible fit for $\alpha_s = 0.1185\pm 0.0025$ with a 
$\chi^2/{\mathrm{d.o.f.}}$ close to unity.

\newpage
\end{document}